\documentclass[letterpaper]{article} 
\usepackage{aaai25}  
\usepackage{times}  
\usepackage{helvet}  
\usepackage{courier}  
\usepackage[hyphens]{url}  
\usepackage{graphicx} 
\usepackage{natbib}  
\usepackage{caption} 
\usepackage{algorithm}
\usepackage{algorithmic}

\let\hat\widehat

\newcommand{\cD}{\mathcal{D}}
\newcommand{\EE}{\mathbb{E}}
\def\given{{\,|\,}}

\usepackage{xcolor}
\usepackage{amssymb}
\usepackage{amsmath}

\usepackage{newfloat}
\usepackage{listings}
\DeclareCaptionStyle{ruled}{labelfont=normalfont,labelsep=colon,strut=off} 
\floatstyle{ruled}
\newfloat{listing}{tb}{lst}{}
\floatname{listing}{Listing}
\nocopyright 

\title{Enterprise Experimentation with Hierarchical Entities}
\author {
    Shan Ba,
    Shilpa Garg,
    Jitendra Agarwal,
    Hanyue Zhao
}
\affiliations{
    LinkedIn Corporation\\
    
    700 E Middlefield Rd,\\
    Mountain View, CA 94043 USA\\
}

\begin{document}

\thispagestyle{plain}

\maketitle

\begin{abstract}
In this paper, we address the challenges in running enterprise experimentation with hierarchical entities and present the methodologies behind the implementation of the Enterprise Experimentation Platform (EEP) at LinkedIn, which plays a pivotal role in delivering an intelligent, scalable, and reliable experimentation experience to optimize performance across all LinkedIn's enterprise products.
We start with an introduction to the hierarchical entity relationships of the enterprise products and how such complex entity structure poses challenges to experimentation.
We then delve into the details of our solutions for EEP including taxonomy based design setup with multiple entities, analysis methodologies in the presence of hierarchical entities, and advanced variance reduction techniques, etc. Recognizing the hierarchical ramping patterns inherent in enterprise experiments, we also propose a two-level Sample Size Ratio Mismatch (SSRM) detection methodology. This approach addresses SSRM at both the randomization unit and analysis unit levels, bolstering the internal validity and trustworthiness of analysis results within EEP. 
In the end, we discuss implementations and examine the business impacts of EEP through practical examples. 
\end{abstract}

%

\section{Introduction} \label{sec_intro}

The LinkedIn ecosystem propels member and customer value through a series of enterprise products, including talent solutions (for job seekers and recruiters), marketing solutions (for advertisers), sales solutions and learning solutions.
The optimization of this value is achieved through the strategic utilization of data-informed decision-making and the integration of A/B testing~\citep{kohavi_tang_xu_2020} for more precise measurement of feature performance across LinkedIn's products.
Enterprise products at LinkedIn used to suffer from inadequate experimentation capabilities due to several challenges associated with the intricate nature of its entity relationships. 

\begin{figure}[h]
  \centering
  \includegraphics[width=0.5\textwidth]{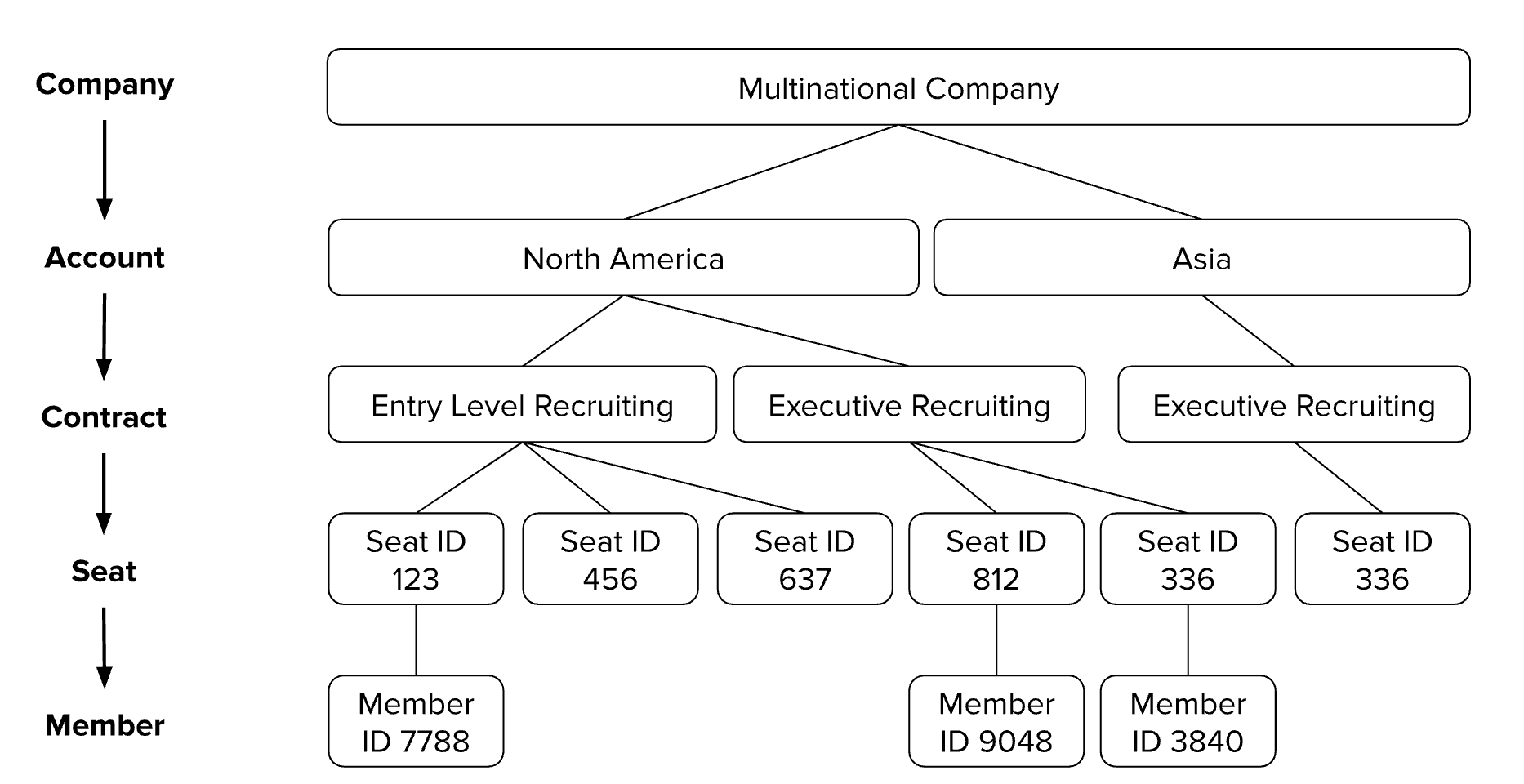}
  \caption{Hierarchical Entity Relationship in LinkedIn Talent Solutions (Recruiter Products).}\label{fig_intro_hierarchy}
\end{figure}

(1) Different from individual consumers, the enterprise customers purchase LinkedIn's products (Recruiter, Sales, Learning) under contract or account entity, and under each contract or account, there are seats or profiles ranging from ten to ten thousands, therefore form the “Hierarchical entity relationships” (Figure~\ref{fig_intro_hierarchy}). 
When we launch a new feature or deramp an existing feature with A/B testing to measure its impact, the enterprise customers often time are very sensitive to such change and require “same account same experience” to ensure all seats (i.e., recruiters, sales representatives, etc.) under the same contract or account get consistent experience during A/B testing. Therefore, in designing enterprise experiments, it is imperative to use “higher order entity” such as contract as the randomization entity, while the success metrics we are interested in are the “lower-order” seat entity metrics.

(2) Enterprise experimentation faces the small sample size and high variance problem. Because the experiment is randomized by contract / account entity and the total number of contracts / accounts is two or more order of magnitude smaller than the total number of seats / members, the enterprise experimentation can have two or more orders of magnitude larger variance than seat / member level experiments (i.e., a common consumer experimentation). Enterprise experimentation also tends to suffer from outliers due to high heterogeneity among accounts / contracts. For instance, a contract with a multinational company customer may encompass 10,000 or more seats, while a contract with a small business customer may include only 2 to 10 seats.

(3) The complex entity relationships in enterprise experiments also greatly complicates the Sample Size Ratio Mismatch (SSRM) issue. SSRM represents the situation where the observed sample size ratio (treatment sample size/control sample size) in the experiment is different from the expected ratio~\citep{Fabijan2019}. A prior analysis revealed that approximately 10\% of triggered analyses at LinkedIn exhibited SSRM~\citep{chen2019}. In order to ensure the internal validity and trustworthiness of the analysis results, SSRM analysis should be included for every experiment~\citep{kohavi_tang_xu_2020}. When SSRM is detected, it signals experiment is bias, rendering metric analyses invalid, and the experiment owner needs to diagnose/fix the issue before interpreting the experiment readout. While SSRM is a well-explored topic in regular member-randomized experiments~\citep{Fabijan2019}, SSRM under the hierarchical entity relationships have not been previously studied in the literature. 
The absence of a mechanism to detect SSRMs in EEP poses the risk that an ineffective treatment may erroneously seem beneficial in the enterprise experiments and be deployed to users.

\section{Taxonomy based Design Setup }\label{sec:taxonomy_setup}

In traditional consumer experiments, the randomization, targeting, and success metric measurement are typically conducted on the same entity (e.g., member ID, guest cookie browser ID, etc.). However, in enterprise experimentation, EEP offers a high degree of flexibility.
Users have the capability to utilize multiple entities for setting up randomization, targeting conditions, and success metric entities, as long as the entities and relationship are compliant with the taxonomy of the business line.

\begin{figure}[h]
  \centering
  \includegraphics[width=0.4\textwidth]{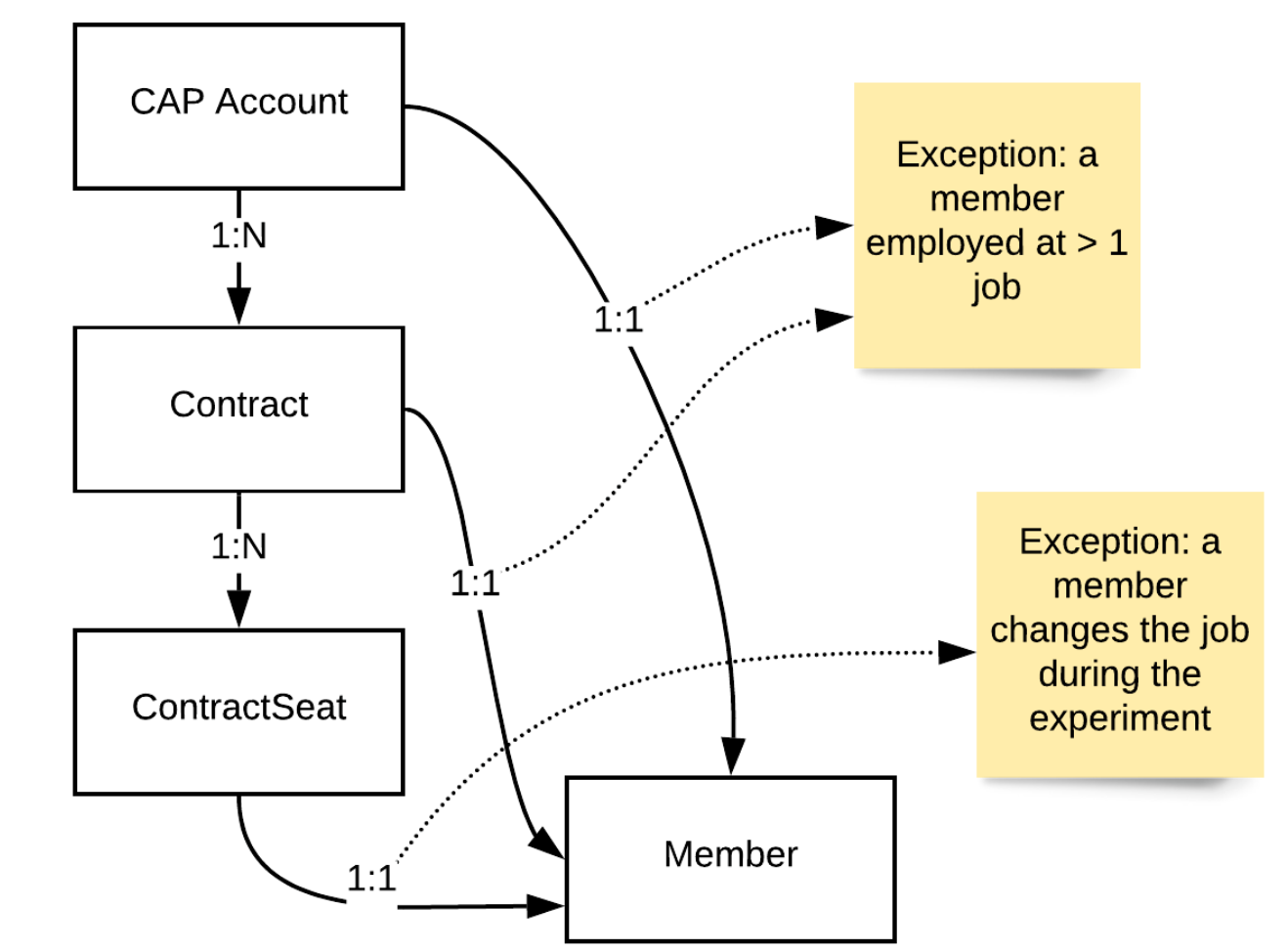}
  \caption{Taxonomy of LinkedIn Talent Solutions.}\label{fig_taxonomy}
\end{figure}

Greater flexibility in the setup comes with a greater complexity. Users at all levels have different knowledge in the complex domain and unrestricted configuration can be error-prone. Therefore we have introduced a formal model of the LinkedIn Enterprise domain, called “taxonomies”, which define entities and types of relationships (Figure~\ref{fig_taxonomy}). Taxonomies are used to limit users’ selections to a corresponding subset of entities and restrict their selections for entities used in A/B metric attribution and randomization.    
For example, users’ test setup can only use the entities included in their corresponding taxonomy (i.e., Talent Solution cannot use advertiser which is a Marketing Solution entity in setup) and the hierarchical ramp must follow a strict 1:N relationship to yield valid A/B testing result (i.e., Sales Solution can randomize at the higher order “contract” entity and measure at the lower order “seat” entity, but cannot do vice versa).

\section{Variance Estimation with Hierarchical Entities} \label{sec:var_estimation}

Enterprise experiments have \emph{misaligned randomization unit and analysis unit}: the experiment needs to be randomized by accounts or contracts to ensure ``same account same experience'', but the success metric for analysis are the ``lower-order'' seat entity metrics. These metrics align with standard practices, facilitate comparisons across experiments, and avoid inconsistencies in the size of ``higher-order'' entities like accounts or contracts, which can vary significantly and lack stability over time.

Due to the misalignment of randomization unit and analysis unit, all metrics of interest in enterprise experiments need to be analyzed as ratio metrics. 
Mathematically, suppose that $n$ contracts $i=1,\dots,n$ are randomly allocated to treated or control groups in an enterprise experiment. Let $Y_i$ represent the revenue of contract $i$ and $N_i$ represent the number of seats in contract $i$, both of which are count metrics. Because contracts match the randomization units in the experiment, they can be assumed to be independent and we can directly calculate $Var(Y)$ and $Var(N)$ using the sample variance formula.
When it comes to revenue per seat (our metric of interest in the enterprise experiment), however, 
the sample variance formula cannot be directly applied to calculate their variance because the seats under each contract are not independent. 
Instead, we need to view the seat-level metric (revenue per seat) as a ``ratio'' derived from two contract-level count metrics, which can be defined as $Z$ below: 
\begin{equation}
Z_{\textrm{treated}} = \frac{  \sum_{i~\textrm{treated}} Y_i}{ \sum_{i~\textrm{treated}} N_i}  
= ~
\frac{\frac{1}{n_t} \sum_{i~\textrm{treated}}Y_i }{\frac{1}{n_t} \sum_{i~\textrm{treated}} N_i}
\end{equation} and
\begin{equation}
Z_{\textrm{control}} = \frac{ \sum_{i~\textrm{control}}Y_i}{ \sum_{i~\textrm{control}} N_i}
= ~
\frac{\frac{1}{n_c} \sum_{i~\textrm{control}}Y_i }{\frac{1}{n_c} \sum_{i~\textrm{control}} N_i},
\end{equation}
Note that $Y_i$ and $N_i$ are aggregated across all contracts (randomization units) in the treatment/control groups first before calculating the ratio. 
Because $\bar{Y}$ and $\bar{N}$ are jointly normal based on the central limit theorem, $Z = \bar{Y} / \bar{N}$ is also normally distributed, whose variance can be calculated by the delta method:
\begin{equation}
Var(Z) = \frac{1}{\bar{N}^2}Var(\bar{Y}) + \frac{\bar{Y}^2}{\bar{N}^4}Var(\bar{N}) - 2 \frac{\bar{Y}}{\bar{N}^3} Cov(\bar{Y}, \bar{N}).
\end{equation}

Before EEP has been made generally available at LinkedIn, a common mistake many analysts made in analyzing enterprise experiment was considering revenue per seat as a count metric $Z_i = Y_i/N_i$ and computing its variance by the sample variance formula
$Var(Z) = \frac{1}{n-1} \sum_{i=1}^n (Z_i - \bar{Z})^2$.
This would lead to incorrect variance estimate because $Z_i$ $(i=1,\dots,n)$ are not independent in an enterprise experiment that was randomized by a higher-order entity such as the contract.

\section{Variance Reduction} \label{sec:var_reduction}

Because enterprise experiments have low sample size and highly heterogeneous experimental entities (i.e., account, contract), it is important to apply variance reduction techniques to ensure that they could have enough statistical power in detecting treatment effects.

\subsection{Basic Variance Reduction}
In EEP, we reduce variance by leveraging covariates that are independent of the treatment but correlated with the experimental outcomes. The default analysis pipeline in EEP uses the CUPED methodology \citep{deng2013improving}, which leverages pre-experiment metrics as the covariates. Suppose $T_i\in\{0,1\}$ represents whether contract $i$ has been randomly assigned into the control or treatment group, $Y_i$ is a count metric at the contract level (e.g., revenue of contract $i$) during the experiment period, $N_i$ is the number of seats triggered in contract $i$ during the experiment period, and $X_i$ and $M_i$ are the corresponding measurements of $Y_i$ and $N_i$ during the pre-experiment period. Compared to the regular difference-in-mean estimator (without variance reduction):
\begin{equation} \label{eq:ratio_diff_in_mean}
\frac{\sum_{T_i=1}Y_i}{\sum_{T_i=1}N_i} -  \frac{\sum_{T_i=0}Y_i}{\sum_{T_i=0}N_i},
\end{equation}
CUPED estimates the treatment effect by:
\begin{equation} \label{eq:ratio_cuped}
\frac{\sum_{T_i=1}Y_i}{\sum_{T_i=1}N_i} -  \frac{\sum_{T_i=0}Y_i}{\sum_{T_i=0}N_i} -  \theta \cdot \Bigg( \frac{\sum_{T_i=1}X_i}{\sum_{T_i=1}M_i} -\frac{\sum_{T_i=0}X_i}{\sum_{T_i=0}M_i} \Bigg),
\end{equation}
where
\begin{equation}
\theta = cov(\frac{\bar{Y}}{\mu_N} - \frac{\mu_Y \bar{N}}{\mu_N^2}, 
\frac{\bar{X}}{\mu_M} - \frac{\mu_X \bar{M}}{\mu_M^2})
/ var(\frac{\bar{X}}{\mu_M} - \frac{\mu_X \bar{M}}{\mu_M^2}).
\end{equation}
Because the pre-experiment metrics $X_i$ and $M_i$ typically have a high correlation with the experimental outcomes $Y_i$ and $N_i$, they can be used to largely remove the pre-existing difference among the experimental entities. The EEP variance reduction pipeline has implemented both the regression adjustment method and the stratification method including outlier capping capabilities. 

\subsection{Advanced Variance Reduction}
In addition to the default pipeline, we have also developed an advanced nonlinear variance reduction solution for EEP which can leverage a large number of covariates (i.e., not just the pre-experiment metrics \citep{deng2013improving} and utilize nonlinear adjustment models  (i.e., extending the linear adjustment method from CUPED to flexible machine learning methods \citep{guo2024, JinBa2023}. Let $X$ denote a rich class of covariates where the pre-experiment metric is included as a special case, and $\hat\mu^Y(.)$ and $\hat\mu^N(.)$ represent some machine learning predictors for $Y$ and $N$ based on $X$.
In order to achieve unbiased variance reduction, it is important to eliminate two types of biases: (1) ``\emph{regressor bias}'' from $\hat\mu(.)$ whose convergence rate could be slower than $n^{-1/2}$ without a well-posed parametric model; (2) ``\emph{double-dipping bias}'' which occurs if the same dataset is used both for model-fitting and for prediction. 
Algorithm~\ref{alg:ratio} describes our proposed variance reduction procedure which introduces de-biasing terms to correct the regressor bias and also employs the cross-fitting technique to remove the ``double-dipping bias''.
The first step is the $K$-fold sample splitting for $\cD = (Y_i,N_i,T_i,X_i)_{i=1}^n$ and then the second step is cross-fitting: for each $k\in[K]$, we use the data $\big\{(X_i,N_i ,Y_i)\colon T_i=1, i\in\cD^{(-k)} \big\}$ to obtain estimators $\hat\mu_1^{Y,(k)}(x)$ for $\EE[Y(1)\given X=x]$ and $\hat\mu_1^{N,(k)}(x)$ for $\EE[N(1)\given X=x]$. 
Likewise, we use $\big\{(X_i,N_i,Y_i)\colon T_i=0, i\in\cD^{(-k)} \big\}$ to obtain $\hat\mu_0^{Y,(k)}(x)$ and $\hat\mu_0^{N,(k)}(x)$. Then, we 
calculate predictions $\hat\mu_w^Y(X_i) = \hat\mu_w^{Y,(k)}(X_i)$ and  $\hat\mu_w^N(X_i) = \hat\mu_w^{N,(k)}(X_i)$ for all $i\in \cD^{(k)}$, $w\in \{0,1\}$. 
Finally, we estimate the treatment effect by
\begin{equation} \label{eq:est_delta}
 \frac{ \sum_{i=1}^n A_i  }{ \sum_{i=1}^n B_i}- \frac{ \sum_{i=1}^n C_i  }{ \sum_{i=1}^n D_i}, 
\end{equation}
where $A_i = \hat\mu_1^Y(X_i ) + \frac{T_i}{\hat p}  \big(Y_i - \hat\mu_1^Y(X_i ) \big)$, 
$B_i =  \hat\mu_1^N(X_i ) + \frac{T_i}{\hat p}  \big(N_i - \hat\mu_1^N(X_i ) \big)$,
$C_i =  \hat\mu_0^Y(X_i ) + \frac{1-T_i}{1-\hat p}  \big(Y_i - \hat\mu_0^Y(X_i ) \big)$,  and 
$D_i =  \hat\mu_0^N(X_i ) + \frac{1-T_i}{1-\hat p}  \big(N_i - \hat\mu_0^N(X_i ) \big)$ for  $\hat p = n_t/n$.  
Compared to the difference-in-mean estimator in (\ref{eq:ratio_diff_in_mean}), the estimator in (\ref{eq:est_delta}) can be viewed as substituting the sample means of the treated and control groups with averages of the fit-and-debias predictions for the potential outcomes of all $n$ units (e.g., contracts). It can be proved that the variance reduction procedure in Algorithm~\ref{alg:ratio} is finite sample unbiased and asymptotically optimal (in the sense of semi-parametric efficiency) among all regular estimators as long as the machine learning estimators are consistent, without any requirement for their convergence rates \citep{JinBa2023}.
In practice, the proposed advanced nonlinear variance reduction methodology can further reduce up to 30\% of variance compared to CUPED by going beyond linearity and incorporating a large number of extra covariates.

\begin{algorithm}[h]
\caption{Advanced variance reduction solution by leveraging flexible nonlinear models with a large number of covariates}\label{alg:ratio}
\begin{algorithmic}[1]
\STATE Input: Dataset $\cD =\{(Y_i,X_i,N_i,T_i\}_{i=1}^{n}$, number of folds $K$.
\STATE Randomly split $\cD$ into $K$ folds $\cD^{(k)}$, $k=1,\dots,K$.
\FOR{$k=1,\dots,K$} 
\STATE Use all $(X_i,N_i,Y_i)$ with $T_i=1$ and $i\notin \cD^{(k)}$ to obtain $\hat\mu_1^{Y,(k)}(x)$ and $\hat\mu_1^{N,(k)}(x)$; 
\STATE Use all $(X_i,N_i,Y_i)$ with $T_i=0$ and $i\notin \cD^{(k)}$ to obtain $\hat\mu_0^{Y,(k)}(x)$ and $\hat\mu_0^{N,(k)}(x)$;
\STATE Compute $\hat\mu_w^Y(X_i) = \hat\mu_w^{Y,(k)}(X_i)$ and  $\hat\mu_w^{N}(X_i) = \hat\mu_w^{N,(k)}(X_i)$ for all $i\in \cD^{(k)}$ and $w\in\{0,1\}$.
\ENDFOR
\STATE Compute the estimator according to~\eqref{eq:est_delta}.
\end{algorithmic}
\end{algorithm}

\section{Two-Level Sample Size Ratio Mismatch} \label{sec:ssrm}

As one of the most important indicators of data quality issues and potential biases in large-scale experimentation, Sample Size Ratio Mismatch (SSRM), a.k.a. Sample Ratio Mismatch (SRM), has been reported by multiple companies as a common occurrence and emphasized for its value as a key guardrail \citep{Kohavi_Longbotham2017, Zhao2016, chen2019, Fabijan2019, kohavi_tang_xu_2020}.
Different from the regular member-level experiments, EEP is characterized by hierarchical ramping patterns, whose SSRM issue becomes more complicated and has not been well studied in the literature. 

By default, EEP is triggered by its analysis unit (individual seats), which is more granular than its randomization unit (the whole contracts). 
Consequently, it is typical that only a portion of the seats within a contract are triggered in each enterprise experiment. It is preferable to exclude dormant seats or inactive members from the experiment analysis as they would only dilute the treatment signal with noise. For instance, following the purchase of a LinkedIn Learning contract by an enterprise customer covering all its employees, only a subset of employees may participate in the online learning course during the experiment period. 
By only triggering a subset of active seats from a contract in each enterprise experiment, EEP effectively filters out noise generated by dormant seats unaffected by the experiment treatment, enhancing sensitivity and experiment power.

\begin{figure}[h]
  \centering
  \includegraphics[width=0.3\textwidth]{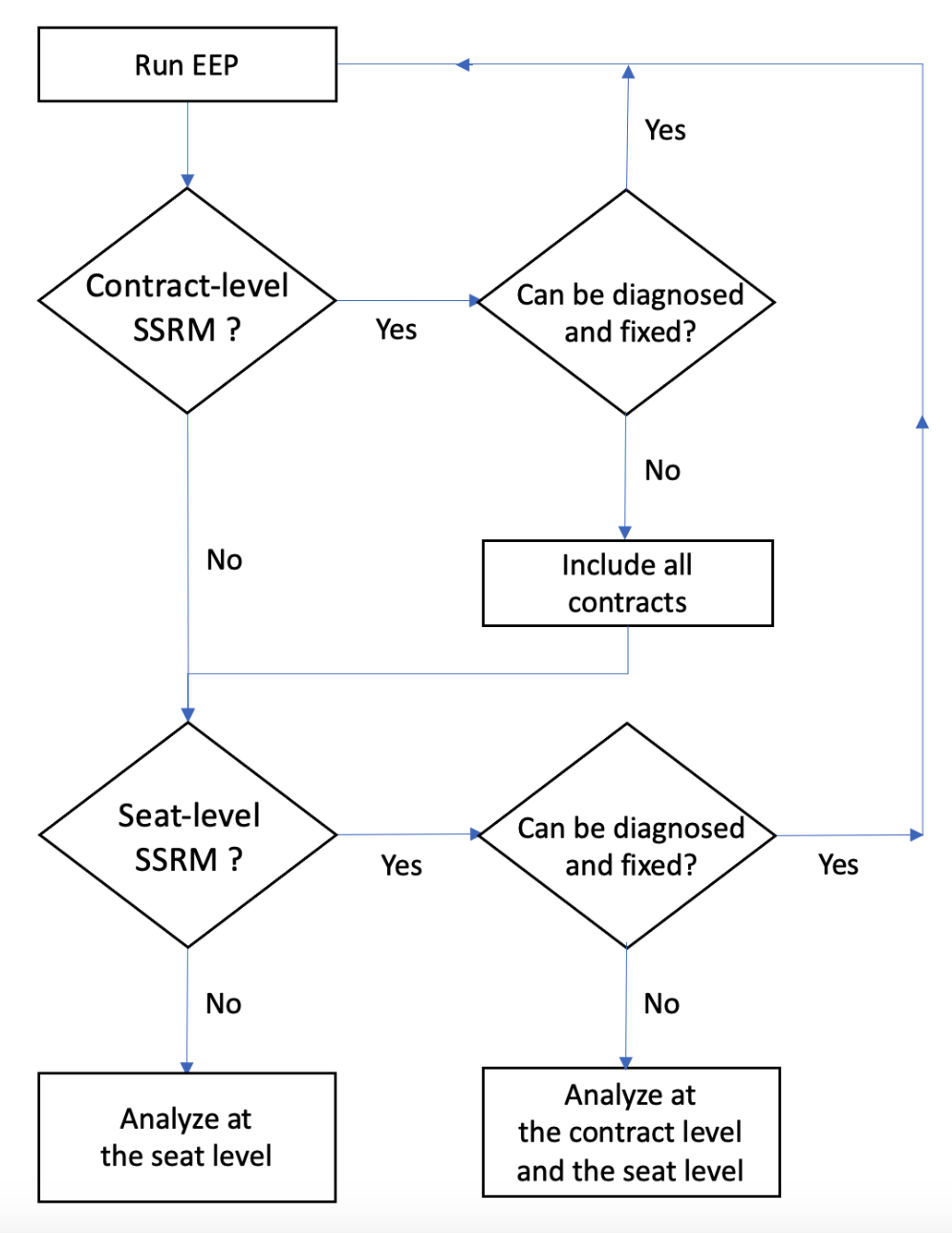}
  \caption{Flowchart for Sample Size Ratio Mismatch (SSRM) detection in EEP.}\label{fig_SSRM}
\end{figure}

SSRM indicates a significant discrepancy between the observed ratio of triggered units across different experiment variants and the expected ratio as per the experiment's design. Previous studies in the literature have primarily addressed SSRM at the randomization unit level.
Recognizing the hierarchical structure described above,  we propose to examine two distinct types of SSRMs within EEP (or more generally, in a cluster-randomized experiment): one at the randomization unit level (contract-level SSRM) and the other at the analysis unit level (seat-level SSRM). Both of them are essential for safeguarding the trustworthiness of the experiment results (Figure~\ref{fig_SSRM}).

\subsection{SSRM at the Randomization Unit Level (Contract-Level SSRM)}

For SSRM at the randomization unit level (contract-level SSRM), the sample size ratio is defined as (number of randomization units in treatment)/(number of randomization units in control). 
The expected sample size ratio for each experiment is directly determined by how we split and allocate online traffic during the experiment configuration. 
For example, if 50\% of traffic is allocated to the treatment group and the other 50\% to the control group, the expected sample size ratio is 1. After running the experiment, the observed sample size ratio should be close to the expected ratio, because the probability of a unit being triggered into the experiment should be independent of any treatment effect. 

SSRM at the randomization unit level (contract-level SSRM) aligns with the conventional SSRM in the literature. Given that randomization units (contracts) can be treated as independent and identically distributed (i.i.d.), we can detect an SSRM at the randomization unit level with a standard chi-square test.
If the observed sample size ratio is statistically different from the expected one, an SSRM alert needs to be fired, which indicates that there is a bias in the experiment setup.

When SSRM at the randomization unit level (contract-level SSRM) occurs, diagnosing its underlying cause follows a similar approach to traditional SSRM diagnosis. Typical root causes for SSRM include residual effects, tracking data issues,  interference with other experiments, and biased experimental design, etc. See \citep{Fabijan2019} and \citep{chen2019} for a comprehensive review on the SSRM root cause diagnosis. 
Once the root cause of SSRM is identified and rectified, the experiment should retake the contract-level SSRM test. If the SSRM root cause cannot be found or is hard to address, we can include all the contracts and seats (instead of only those triggered ones) in the analysis. While this may dilute the signal, it can effectively mitigate bias stemming from SSRM. Only upon successfully passing the contract-level SSRM test can we proceed to evaluate SSRM at the analysis unit level.

\subsection{SSRM at the Analysis Unit Level (Seat-Level SSRM)}

Addressing seat-level SSRM is more relevant for ensuring the trustworthiness of EEP results compared to addressing SSRM at the contract level, because EEP is triggered by individual seats rather than by entire contracts and the EEP analysis is also based on metrics per seat rather than per contract.

Consider a seat-level SSRM scenario where a bad treatment feature substantially reduces the number of active users (seats) within each contract. Failure to detect the seat-level SSRM could erroneously favor the ineffective treatment variant. This is because the treatment variant may exhibit a more favorable revenue-per-user ratio than the control variant, given that the remaining users within the contracts are likely the most active ones.
Unfortunately, in this scenario, the traditional SSRM detection solution at the randomization unit level (contract-level SSRM) would not be effective. This is because a contract would be triggered into the experiment as long as at least one of its seats is active/triggered. Although the treatment variant has a considerable decrease in active seats within the contracts, its overall number of contracts triggered in the experiment may not be significantly different from the control variant. In general, the seat-level SSRM is more likely to occur than the contract-level SSRM. 
To the best of our knowledge, the detection of SSRM at the analysis unit level for a cluster-randomized experiment has not been presented in the literature.

Let $Y/N$ represent a ratio metric in EEP, where $Y$ represents the ``metric per contract'' and $N$ is the ``number of active seats per contract''. 
A crucial assumption on this ratio metric analysis is that
the denominator $N$ remains unchanged by the treatment, which is also referred to as the \emph{stable denominator assumption} in the literature \cite{deng2017trustworthy}.
In EEP, the denominator $N$ essentially serves as a normalization factor, standardizing changes in the numerator metric $Y$. 
If the treatment also alters $N$, the ratio metric $Y/N$ becomes unsuitable for decision-making, as it becomes challenging to interpret whether a change in the ratio value is beneficial or detrimental.
For instance, although a higher revenue-per-seat ratio typically suggests a favorable treatment, a detrimental treatment which substantially decreases the number of active seats in contracts may also yield a higher ratio, as the remaining few seats are likely to be the most active seats.
Thus, unless the number of active seats in contracts remains stable during the experiment, the revenue-per-seat ratio lacks interpretability and may lead to erroneous conclusions. 
The testing of analysis-unit-level SSRM in EEP aims to ensure the stability of the denominator metric $N$, enabling meaningful conclusions based on the ratio metric $Y/N$.

Compared to the previous randomization-unit-level (contract-level) SSRM, detecting the analysis-unit-level (seat-level) SSRM presents two unique challenges. Firstly, the ``expected sample size ratio'' is unknown at the analysis unit level (seat level). 
As discussed in the previous section, it is easy to know the ``expected sample size ratio'' at the randomization unit level, as it is directly determined by the allocation of online traffic. However, because each randomization unit (contract) contains a different number of analysis units (seats), the sample size ratio at the analysis unit level can be quite different due to the high heterogeneity in the sizes of the randomization units.
Secondly, unlike the randomization units (contracts) in the experiment, the analysis units (seats) are no longer independent. Consequently, the classic chi-square test cannot be used to directly compare the number of analysis units in the treatment group versus the control group and the detection method for SSRM at the analysis unit level must diverge from the existing approaches at the randomization unit level.

Our proposed solution to detect analysis-unit-level (seat-level) SSRM is summarized in Algorithm~\ref{alg:ssrm}. 
It is important to ensure that there is no contract-level SSRM before running this procedure.
In step 2, for each triggered contract in the experiment, $N^{pre}$ represents the pre-experiment number of active seats, where the pre-experiment period needs to free from the treatment effect.  
Step 3 involves using $N_i^{pre}$ as a baseline to adjust for pre-existing size differences among contracts, which is crucial due to the high heterogeneity in contract size in EEP. By adjusting for this baseline variation, we can largely mitigate pre-experiment difference in average contract size between the treatment and control groups and also enhance the sensitivity of the SSRM test.  
The coefficient $\theta$ in step 3 equals to the ordinary least square solution of regressing (centered) $N^{triggered}$ on (centered) $N^{pre}$. This choice of $\theta$ minimizes $var(D)$, which is similar to the CUPED estimator.  
Under the traditional SSRM detection framework, $N^{triggered}$ can be viewed as the ``observed sample size'' in the experiment and  $N^{pre}$ acts as a surrogate for the unknown ``expected sample size'' at the seat level.
Because the test for seat-level SSRM in step 4 is conducted at the randomization unit (contract) level,  it no longer violates the independence assumption of the t test. 

\begin{algorithm}[h]
\caption{Detection of analysis-unit-level (seat-level) SSRM} \label{alg:ssrm}
\begin{algorithmic}[1]
\STATE Obtain the list of contracts that were triggered in the experiment (assuming no contract-level SSRM).
\STATE Compute the contract-level metrics $N^{triggered}$ and $N^{pre}$ for each triggered contract.  ($N^{triggered}$ represents the number of triggered seats per contract in the experiment period, and $N^{pre}$ represents the number of active seats per contract in the pre-experiment period. )  
\STATE For each triggered contract $i$, compute $$D_i = N_i^{triggered} - \theta(N_i^{pre}  - E(N^{pre})),$$ where  
$\theta = cov(N^{triggered}, N^{pre})/var(N^{pre})$.
\STATE Run a two-sample t test to compare whether the mean of $D_i$ is significantly different between the treatment group and the control group. If the difference is significant, fire a seat-level SSRM alert. 
\end{algorithmic}
\end{algorithm}

When seat-level SSRM occurs, the interpretability of the ratio metric $Y/N$ is compromised, necessitating a diagnostic process akin to debugging traditional SSRM in member-randomized experiments \citep{Fabijan2019, chen2019}. Upon identifying and rectifying the root cause of this seat-level SSRM, it is advisable to rerun both contract-level and seat-level SSRM tests before proceeding with analysis based on the ratio metric $Y/N$. However, if the cause of seat-level SSRM remains elusive or cannot be resolved (while no contract-level SSRM is present), we recommend analyzing the numerator metric $Y$ and the denominator metric $N$ of the original ratio metric $Y/N$ separately. In other words, alongside the original seat-level analysis using the ratio $Y/N$, contract-level analyses based on metrics $Y$ and $N$ should also be conducted separately, employing CUPED adjustment to mitigate pre-existing heterogeneity among contracts. These contract-level analyses remain valid, free from seat-level SSRM bias, and alleviate the interpretational challenges associated with the ratio metric $Y/N$. Subsequently, conclusions can be drawn by collectively interpreting the outcomes of the three tests based on $Y$, $N$, and $Y/N$.

\section{Implementations} \label{sec:implementation}

\subsection{Test Setup Flow}

\begin{figure}[h]
  \centering
  \includegraphics[width=0.5\textwidth]{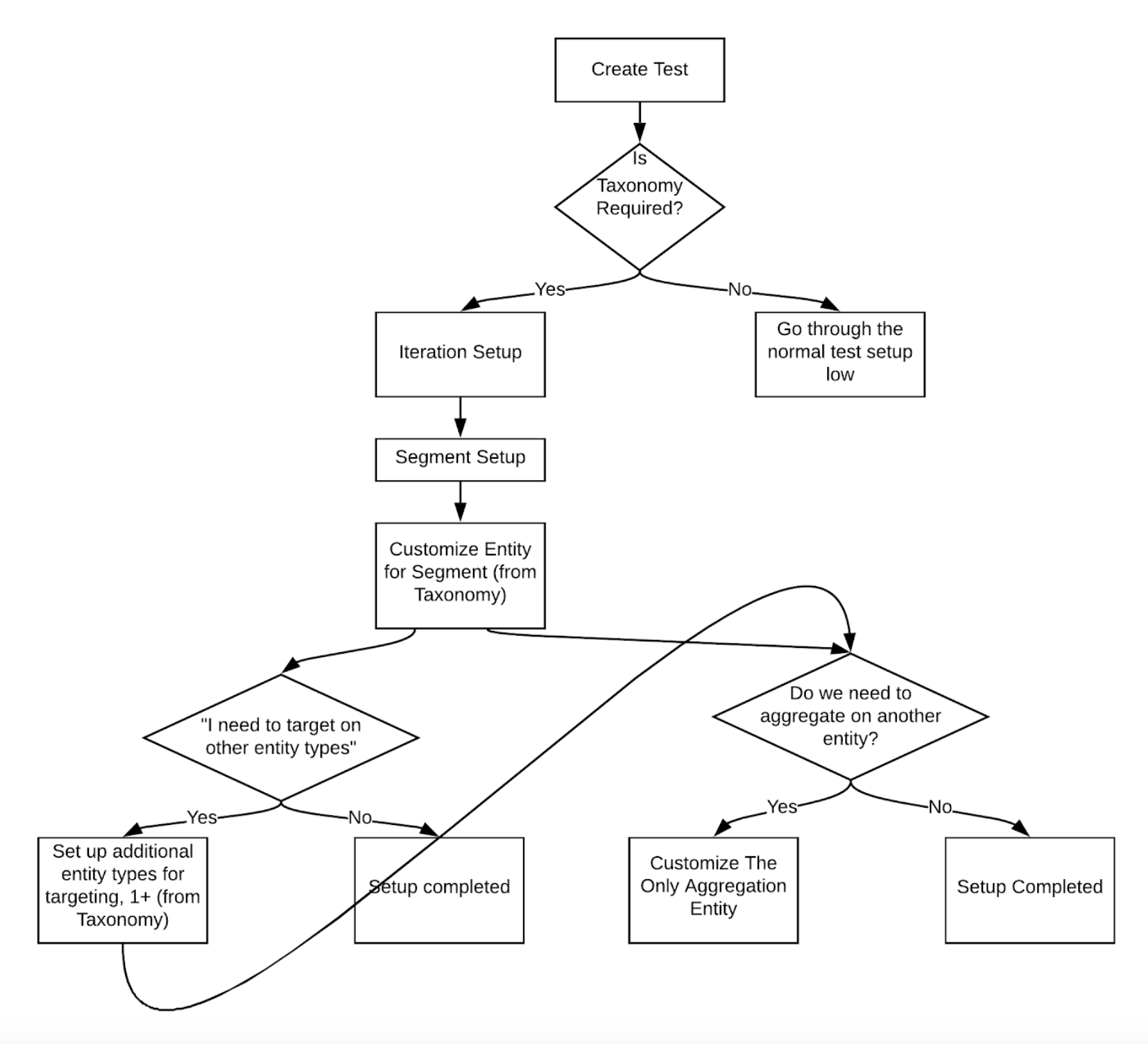}
  \caption{Illustration of EEP test setup flow.}\label{fig_test_setup}
\end{figure}

In EEP, the experiment setup process involves enabling the use of multiple entities for randomization, targeting conditions, and reporting measurement entities, all while ensuring compliance with the business taxonomy. 
To accommodate these requirements, the implementation of EEP has expanded the normal test setup flow of LinkedIn's central experimentation platform as shown in (Figure~\ref{fig_test_setup}).
This expanded setup flow includes selecting either a single entity or a predefined entity taxonomy during test setup. Users can choose entities to target from a predefined set or directly from the Taxonomy dropdown list, which automatically populates entities allowed by the taxonomy. Additionally, users can customize targeting or measurement entity types on the segment setup page, where they can configure segments to modify targeting, aggregation, and reporting entity units. These options are predefined in the taxonomy and dynamically restricted based on the taxonomy definition.

\subsection{Enhanced Tracking Data and Analysis Flow}

\begin{figure}[h]
  \centering
  \includegraphics[width=0.5\textwidth]{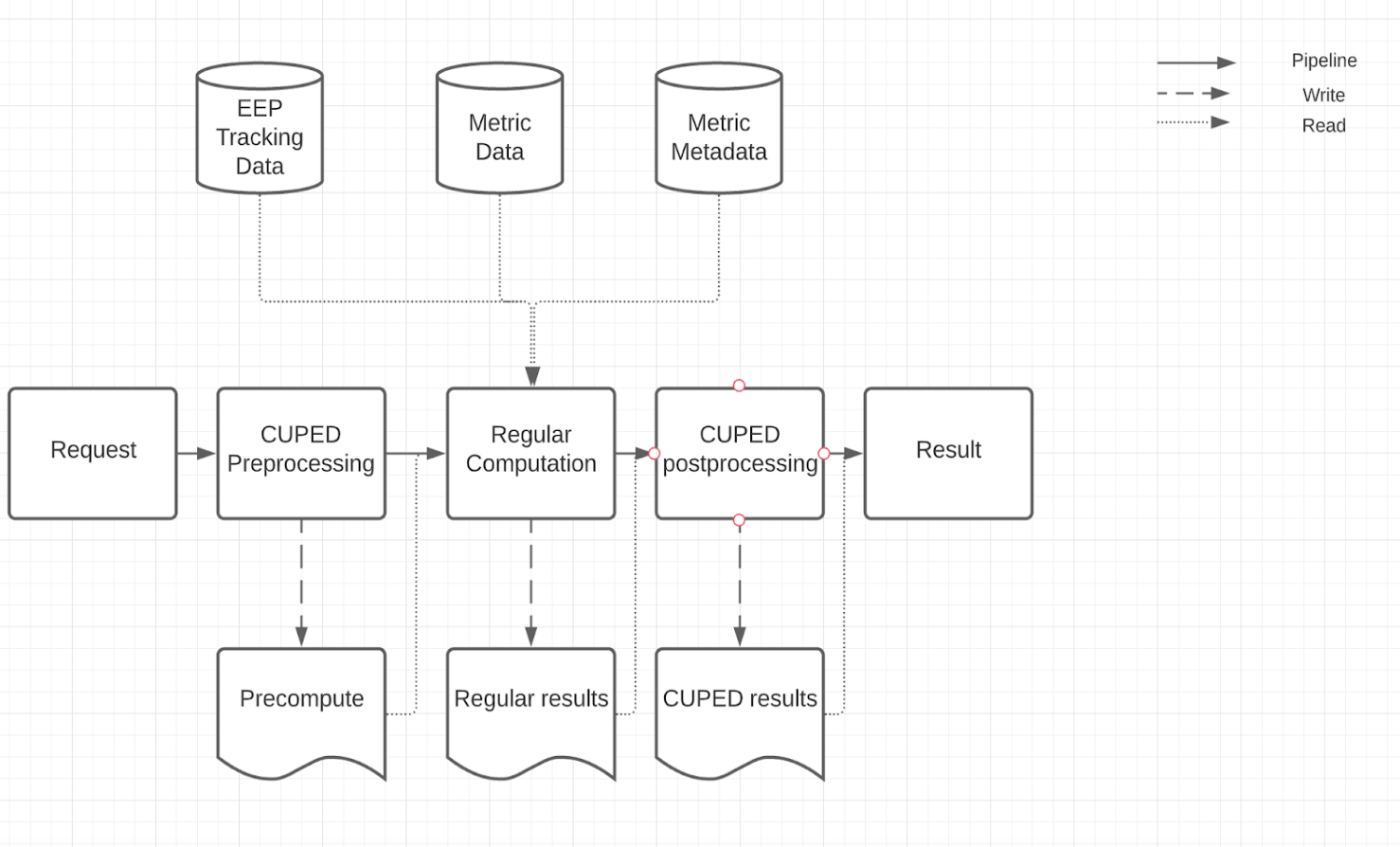}
  \caption{Computation of EEP reports.}\label{fig_computation_of_EEP_reports}
\end{figure}

To generate EEP reports effectively, it is crucial to capture all pertinent entities emitted in the tracking data. To facilitate this, updates have been made to the tracking data schema, incorporating both randomization and analysis entities. This schema encompasses key components such as the test key, experiment ID, segment ID, variant, aggregation URN (randomization unit), join URN (analysis unit), and others.  Enhanced tracking data is combined with metrics data to compute results, as illustrated in Figure~\ref{fig_computation_of_EEP_reports}. Alongside regular results, which include mean, variance, lift, p-value, etc., a symmetric computation flow has been introduced, focusing on the same triggered population but computing results (lift, p-value, etc.) during a pre-experiment period. This approach encompasses A/A results, regular A/B results, and variance reduction (CUPED) results. Outliers are determined based on a specified threshold (e.g., 99th quantile) and their values are capped. All outlier data is centrally stored for user access and consumption.

\subsection{Guardrails}

EEP allows for a great degree of flexibility for  experiment owners, but that also opens up doors for a wide range of experiment-setup issues. Incorrectly configured ramps have the potential to affect experimentation velocity, developer productivity due to increased troubleshooting, and ultimately impact customers' ability to make data-driven decisions for their enterprise features. To maintain simplicity and ease of configuration while safeguarding users from setup issues, several protective measures have been implemented in EEP. For instance, if the experiment is setup incorrectly or the experiment-evaluation code is written incorrectly, it could lead to null or zero tracking data. Email notifications are set up to proactively inform experiment owners of such issues and provide pointers to help them investigate and aid in troubleshooting. The proactive alert signals issues at the early stage, draw attention and greatly help on experimentation velocity and reduce customer experience friction.
Additionally, the accurate deduped triggered sample size data broken down by randomization and analysis units is displayed in the UI. This helps users to quickly assess and catch issues in over/under-triggering, dilution and mistargeting. 
Furthermore, users are alerted if SSRM is detected, ensuring the trustworthiness of reports and enhancing the quality and robustness of insights and decision-making processes.

\section{Results} \label{sec:example}

Since its launch, EEP has garnered widespread adoption across LinkedIn's enterprise business lines and resulted in significant business impacts.

\subsection{Enable and Scale Up Measurements}

Before the implementation of EEP, enterprise-facing feature or infrastructure changes were rolled out to our customers in quarterly batches (known as Quarterly Product Releases) without A/B testing. While we diligently monitored metric changes before and after each release and collected qualitative feedback from our customers, our enterprise business line lacked robust measurement and data-driven insights necessary for optimal decision-making. 
With the introduction of EEP, we have transformed the concept of ``A/B testing being impossible'' into a feasible reality for enterprise products. 

Since its launch, the number of enterprise experiments running on the EEP platform has rapidly scaled: starting with approximately 10 experiments during the pilot phase, the number escalated to over 100 per quarter during the beta phase, surpassed 500 per quarter post the General Availability of EEP, and currently maintains a pace of over 1000 experiments (testing around 300 unique new product features) per quarter. The typical run time of each experiment ranges from 2 to 4 weeks.

\begin{figure}[h]
  \centering
  \includegraphics[width=0.5\textwidth]{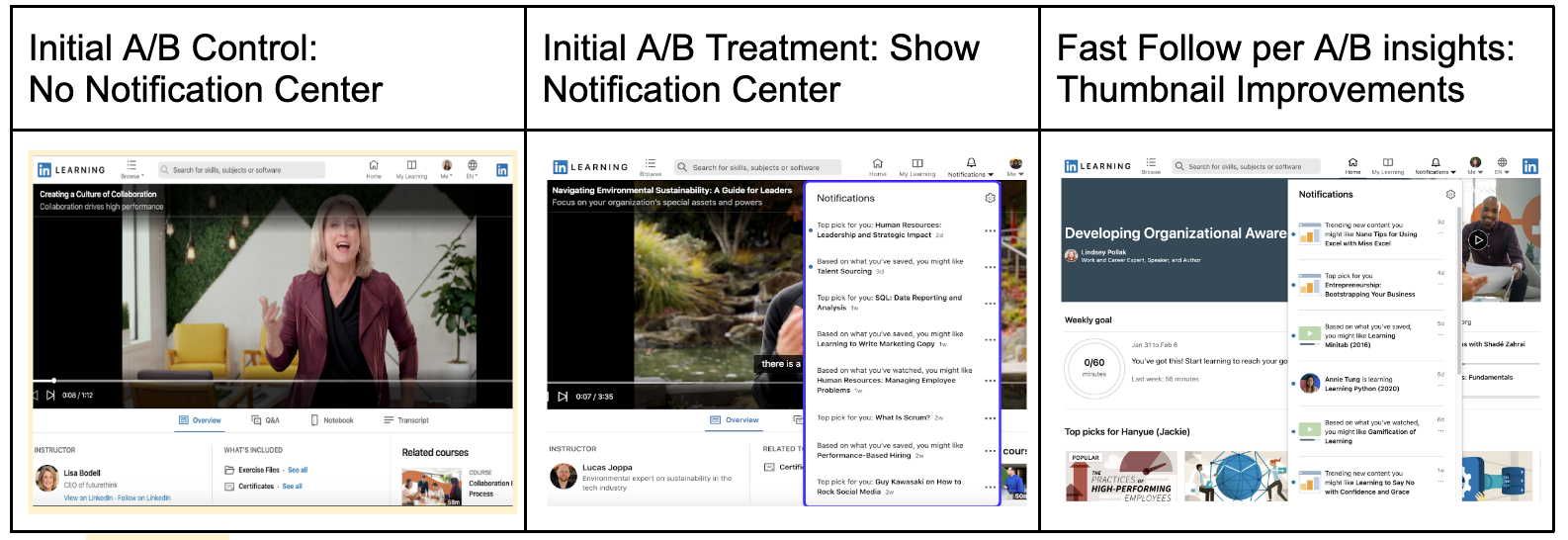}
  \caption{Improve LinkedIn learning notification center via robust A/B tests using EEP.}\label{fig_example_learning}
\end{figure}

Figure~\ref{fig_example_learning} illustrates an instance where EEP's hierarchical ramping facilitates randomization at the enterprise account level, ensuring a consistent experience for all enterprise learners under one account over time. By measuring learner engagement metrics at the individual enterprise profile level, we gain valuable insights into how learners respond to the new notification center. This enables us to continually optimize the user experience, such as improving notification thumbnails, based on insights learned from the previous rounds of A/B testing. 
Without EEP, the release of a feature would lack quantitative measures to gauge its impact on enterprise engagement and retention, hindering our ability to implement necessary improvements.

\subsection{Bolster Trustworthiness through SSRM Guardrails}

EEP has improved the quality of readouts by implementing SSRM guardrail monitoring and reducing manual analysis errors. 
At an aggregated level, we have found approximately 7\% of tests suffer from SSRMs. Our observations include:

(1) The existing single-level SSRM detection method (at the contract level) failed to identify any issues within EEP. This is because a contract is triggered into the experiment if at least one of its seats is triggered. In most cases, contracts containing zero triggered seats were rare, and nearly all contracts were triggered regardless.

(2) The seat-level SSRM (under the proposed two-level SSRM solution) should be the focus for SSRM detection within EEP. Space limitations prevent us from discussing its root cause diagnosis (e.g., due to residual effects, etc.).

\subsection{Expedite Experimentation Velocity and Productivity}

\begin{figure}[h]
  \centering
  \includegraphics[width=0.5\textwidth]{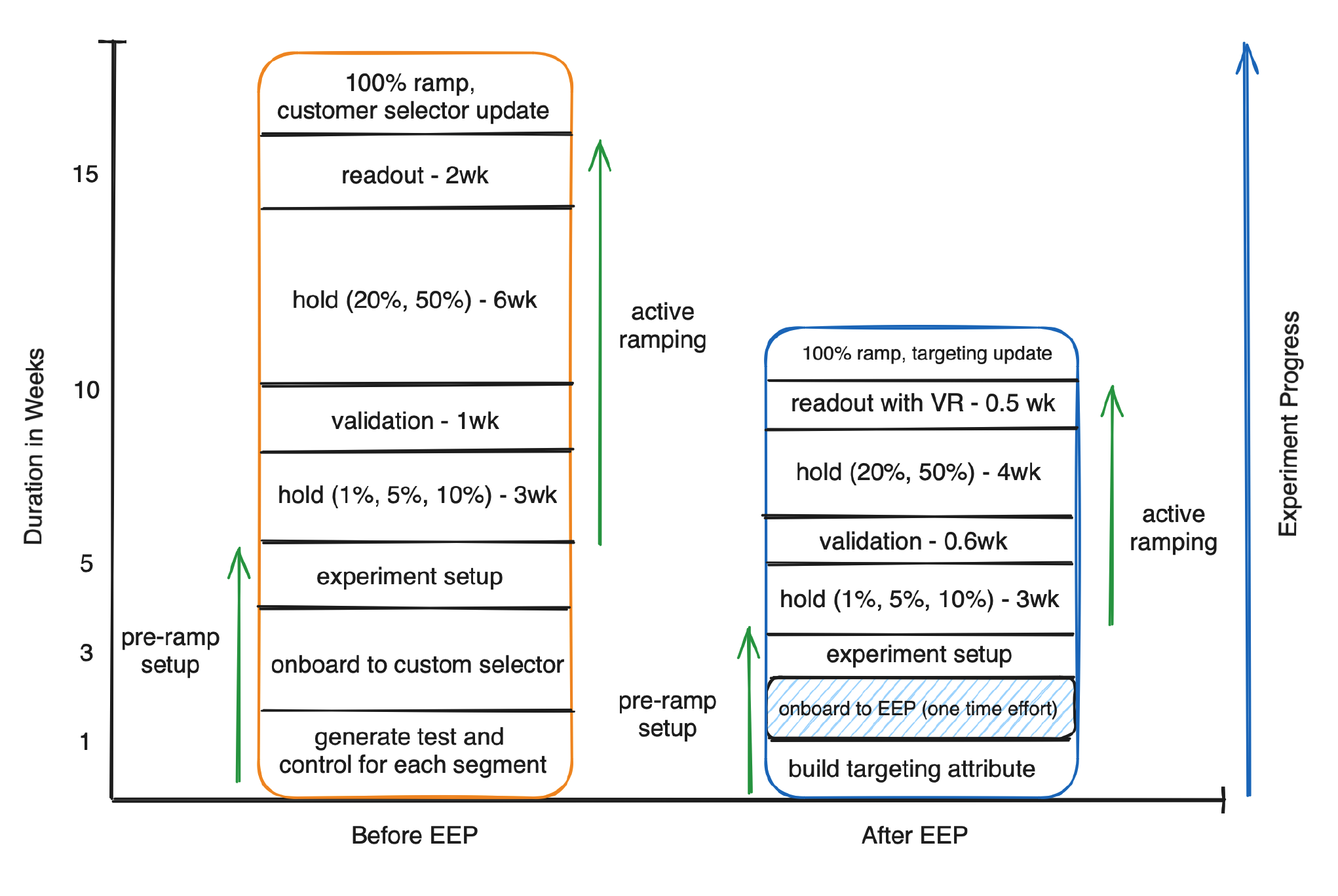}
  \caption{Experimentation speed and effort comparisons with/without EEP.}\label{fig_example_comparison2}
\end{figure}

With EEP, the end-to-end clock time required to leverage online testing for evaluating new enterprise features has significantly improved.
Prior to EEP, the engineering team had to manually generate test / control contracts for each segment, implement workarounds to target contracts and seats correctly, manually check quality guardrails such as SSRM, and execute error-prone manual scripts for compute readouts with variance reduction procedures. With EEP, all these tasks are end-to-end automated by the platform including the advanced variance reduction solution, resulting in a 50\% reduction in efforts and a 2-week reduction in clock time per experiment (Figure~\ref{fig_example_comparison2}). 
Additionally, EEP has streamlined the post-review and quality check process, and introduced a user-friendly readout report UI that allows Product Managers to self-serve, thereby expediting the business decision-making process.

\section{Conclusions} \label{sec:conclusions}

In conclusion, our work on the Enterprise Experimentation Platform (EEP) at LinkedIn has addressed critical challenges posed by the hierarchical entity relationships within enterprise products. By introducing a taxonomy-based design setup, refining analysis methodologies, and implementing advanced variance reduction techniques, EEP stands as a robust framework for conducting intelligent, scalable, and trustworthy experimentation. The proposed two-level Sample Size Ratio Mismatch (SSRM) detection methodology, operating at both randomization unit and analysis unit levels, further enhances the platform's capability to ensure internal validity and the reliability of analysis results. 

\section{Acknowledgments}
We would like to thank our current and former colleagues who have contributed to the realization of designing and building the overall EEP solution. Thanks to 
Alexander Ivaniuk, Weitao Duan, Min Liu, Justin Marsh, Juanyan Li, Ying Jin, Chunzhe Zhang, Wentao Su and special thanks to our leaders for their supports: Ya Xu, Kapil Surlaker, Vish Balasubramanian, Kuo-Ning Huang, Sofus Macskassy, Parvez Ahammad and Robert Kyle.

\bibliography{Reference}

\end{document}